\documentclass[twocolumn,superscriptaddress,aps,longbibliography]{revtex4-1}
\usepackage{amsmath}
\usepackage{amssymb}
\usepackage{bm}
\usepackage{braket}
\usepackage{color}
\usepackage{varwidth}
\usepackage{dcolumn}
\usepackage[breaklinks,colorlinks=true,linkcolor=blue,urlcolor=cyan,citecolor=blue]{hyperref}
\usepackage{graphicx}
\usepackage{algorithm}
\usepackage{algorithmicx}
\usepackage{algpseudocode}
\algdef{SE}[DOWHILE]{Do}{doWhile}{\algorithmicdo}[1]{\algorithmicwhile\ #1}%

\begin{document}

\title{Lifted directed-worm algorithm}

\author{Hidemaro Suwa}
\affiliation{Department of Physics, The University of Tokyo, Tokyo 113-0033, Japan}

\begin{abstract}
Nonreversible Markov chains can outperform reversible chains in the Markov chain Monte Carlo method.
Lifting is a versatile approach to introducing net stochastic flow in state space and constructing a nonreversible Markov chain.
We present here an application of the lifting technique to the directed-worm algorithm.
The transition probability of the worm update is optimized using the geometric allocation approach; the worm backscattering probability is minimized, and the stochastic flow breaking the detailed balance is maximized.
We demonstrate the performance improvement over the previous worm and cluster algorithms for the four-dimensional hypercubic lattice Ising model.
The sampling efficiency of the present algorithm is approximately 80, 5, and 1.7 times as high as those of the standard worm algorithm, the Wolff cluster algorithm, and the previous lifted worm algorithm, respectively.
We estimate the dynamic critical exponent of the hypercubic lattice Ising model to be $z \approx 0$ in the worm and the Wolff cluster updates.
The lifted version of the directed-worm algorithm can be applied to a variety of quantum systems as well as classical systems.
\end{abstract}

\date{\today}
\maketitle
\section{Introduction}
Many physical models, such as the Ising model, can be mapped to configurations of bond variables, also called dimers \cite{Prokof'evS2001}.
The physical bond configurations are restricted by constraints on the state space, and it is nontrivial to sample bond configurations efficiently under the constraint.
The worm algorithm has an advantage in sampling configurations under constraints on state space, and it is one of the most efficient algorithms for various classical and quantum systems \cite{Prokof'evST1998,Prokof'evS2001,BoninsegniPS2006}.
Perhaps surprisingly, the worm algorithm significantly alleviates critical slowing down and reduces the dynamic critical exponent. 
For example, the dynamic critical exponent of the cubic lattice Ising model is $z\approx 0.27$ in the worm update \cite{Suwa2021}, while it is $z \approx 0.46$ \cite{Ossola2004} in the Swendsen-Wang cluster update \cite{SwendsenW1987}.
The worm algorithm has been applied to a wide variety of physical models, such as the $| \phi|^4$ model \cite{Prokof'evS2001}, the Potts model \cite{MercadoEG2012}, the O($n$) loop model \cite{JankeNS2010,LiuDG2011}, frustrated Ising models \cite{RakalaD2017}, spin glasses \cite{Wang2005}, lattice QCD \cite{AdamsC2003}, quantum spins \cite{SyljuasenS2002,GubernatisKW2016,YasudaST2015}, bosons \cite{BoninsegniPS2006,WesselATB2004,SuwaT2015}, and fermions \cite{BurovskiPST2006,GunackerWRHSH2016}.
Thus, it is crucial to improve the worm algorithm and push the limit of the Monte Carlo method for physical models.

The heart of the worm algorithm is to enlarge the state space to those with kinks that break the constraint. 
A nonlocal update in the original state space, otherwise hard to perform, can be achieved as a series of local updates in the enlarged state space. 
In the worm update, the Monte Carlo dynamics can be viewed as a diffusion process of the kinks. 
Because the configuration is updated by diffusion of the kink, a higher diffusivity is expected to yield a higher sampling efficiency. 
Here, a bottleneck of the algorithm is the worm backscattering process, in which the kink traces back the path. 
Because the backscattering, a rejection process in the Markov chain Monte Carlo (MCMC) method, naturally lowers the diffusivity, it is desirable to minimize the worm backscattering probability for efficient sampling \cite{SuwaT2010}.  

As improved versions of the worm algorithm, the directed loop (or the directed worm) algorithm for quantum systems \cite{SyljuasenS2002} and an extended version for classical systems were proposed \cite{Suwa2021}.
In classical cases, the kinks in the directed-worm algorithm are located on bonds of a lattice instead of sites, in contrast to the conventional worm algorithm \cite{Prokof'evS2001}.
The probability of worm scattering can be easily optimized using the geometric allocation approach \cite{SuwaT2010, TodoS2013, Suwa2014}, and the backscattering probability is reduced to zero in many cases. 
The directed-worm algorithm enhances the diffusivity of the kink and significantly improves sampling efficiency.

In the meantime, it has been actively discussed that nonreversible Markov chains can outperform reversible chains \cite{DiaconisHN2000,ChenLP1999}.
It was mathematically proved that adding perturbative stochastic flows breaking detailed balance, or reversibility, always leads to faster distribution convergence \cite{HwangHS2005,IchikiO2013}.
Lifting is a versatile approach to constructing a nonreversible Markov chain, and it has been applied to various physical systems \cite{Vucelja2016}.
The idea of lifting is to enlarge the state space and introduce stochastic flows in the enlarged state space.
The lifted nonreversible Markov chain can achieve up to a square root improvement in the convergence time to the steady state \cite{TuritsynCV2011,FernandesW2011,SakaiH2013}.
Although it is challenging to reach square root improvement in general cases, lifted nonreversible chains can achieve a significant variance reduction in many cases.
A variety of efficient Monte Carlo methods have been developed based on the concept of lifting, such as the event-chain Monte Carlo method \cite{BernardKW2009,MichelKK2014}.
In particular, lifted versions of the worm algorithm \cite{ElciGDNGD2018} and self-avoiding walks \cite{HuCD2017,ZhaoV2022} were recently proposed and shown to be much more efficient than the original reversible algorithm.
The improvement factor is especially large in high dimensions.

In the present paper, we show a lifted version of the directed-worm algorithm to improve the worm algorithm further.
The transition probability is optimized using the geometric allocation approach to minimize backscattering probability and maximize net stochastic flow.
Reviewing the bond representation of the Ising model and the directed-worm algorithm in Sec.~\ref{sec:model}, we provide an optimized probability set for the Ising model on the $d$–dimensional torus in Sec.~\ref{sec:ldwa}.
Investigating the statistical errors of the energy and the magnetic susceptibility at the critical temperature of the four-dimensional hypercubic lattice Ising model, we demonstrate a significant variance reduction in Sec.~\ref{sec:result}: the sampling efficiency of the present worm algorithm is approximately 80, 5, and 1.7 times those of the standard worm algorithm, the Wolff algorithm, and the previous lifted worm algorithm, respectively.
The dynamic critical exponent is estimated to be $z \approx 0$ in the worm and the Wolff algorithms. 
The present paper is summarized with discussions in Sec.~\ref{sec:con}

\section{Model}
\label{sec:model}
The worm algorithm for classical models was originally proposed by Prokof’ev and Svistunov \cite{Prokof'evS2001}, which we call the P-S worm hereafter.
We adopt the same representation of the partition function of the ferromagnetic Ising model on a bipartite lattice. 
The model is represented by $-\beta H = K \sum_{\langle ij \rangle} \sigma_i \sigma_j$, where $\beta=1/T$ is the inverse temperature, $H$ is the Hamiltonian, $\sigma_i=\pm 1$ is the Ising spin at site $i$ of a lattice, and $\langle i,j \rangle$ runs over all the pairs of nearest-neighbor sites.
Using the identity $e^{K \sigma_i \sigma_j}= \cosh(K) \sum_{n_b=0,1}[ \sigma_i \sigma_j \tanh K]^{n_b}$, the partition function is represented as follows:
\begin{eqnarray}
  Z&=&\sum_{\sigma_i = \pm 1} e^{K \sum_{\langle ij \rangle} \sigma_i \sigma_j} \nonumber\\
  &=&\sum_{\sigma_i = \pm 1} \prod_{b=\langle ij \rangle} e^{K \sigma_i \sigma_j} \nonumber \\
  &=& \sum_{\sigma_i= \pm 1} \prod_{b=\langle ij \rangle}  \cosh(K) \sum_{n_b=0,1} [\sigma_i \sigma_j \tanh K]^{n_b} \nonumber \\
  &=& 2^{N} [\cosh K]^{N_b} \sum_{\{n_b\} \in {\rm loop}} [\tanh K]^{\ell},
  \label{eq:Z}
\end{eqnarray}
where $n_b$ denotes the bond variable on bond $b$, $\ell \equiv \sum_b n_b$, and $N$ and $N_b$ are the total numbers of sites and bonds of a lattice, respectively. 
The sum in the last line is taken over all the bond configurations that form loops of the activated bonds ($n_b=1$). Integrating out the spin variables first removes all the bond configurations containing an open string of activated bonds.

The worm algorithm aims to sample bond configurations efficiently under the loop constraint. 
In the worm algorithm, kinks that break the constraint are inserted into the system, and the kink random walk updates the configuration.
A serious bottleneck of the algorithm is worm backscattering, in which the kink traced the path backwards. 
As the backscattering lowers the diffusivity of the kink and the sampling efficiency, it is desirable to minimize the worm backscattering probability for efficient simulations \cite{SuwaT2010, SuwaT2012}.

A directed version of the worm algorithm \cite{Suwa2021} was recently proposed to avoid backscattering.
In contrast to the standard worm algorithm, the kinks are located on bonds and move from bond to bond stochastically, which we call worm scattering hereafter.
The kink is assumed to be located at the center of a bond; the bond variable takes $0, \frac{1}{2},$ or $1$ accordingly. 
The kink possesses the moving direction and keeps it forward after each scattering. 

\section{Lifted directed-worm algorithm}
\label{sec:ldwa}
\subsection{Lifting on the energy axis}
\label{sec:lifting}
We apply lifting to the directed-worm algorithm to further improve the worm algorithm.
In the lifting technique, the state space is enlarged by introducing lifted variables, and net stochastic flows are induced in the enlarged space \cite{DiaconisHN2000,ChenLP1999,Vucelja2016}.
As relevant applications, lifted versions of the Berretti–Sokal (B-S) algorithm \cite{BerrettiS1985} were recently proposed for the Ising model \cite{ElciGDNGD2018} and self-avoiding walks \cite{HuCD2017,ZhaoV2022}. 
In the original B-S algorithm, one randomly chooses to increase $n_b$ or decrease $n_b$ in each local worm update.
When increasing (decreasing) $n_b$, a state with increased (decreased) $n_b$ is randomly chosen and accepted using the standard Metropolis algorithm.
In the lifted B-S algorithm, one keeps the label, $+$ corresponding to the $n_b$ increasing mode and $-$ corresponding to the $n_b$ decreasing mode, and flips it ($+ \leftrightarrow -$ ) only when the proposed state is rejected by the Metropolis filter.
Thus, the state space is enlarged by introducing the lifted variable $\sigma=\pm$.
Because the number of activated bonds $\ell=\sum_b n_b$ corresponds to the energy of the original spin system, as represented in Eq.~(\ref{eq:e}), this approach introduces net stochastic flows on the energy axis.
The lifted B-S worm algorithms significantly improve sampling efficiency compared to the original B-S worm algorithm.

\begin{algorithm}[H]
  \caption{Lifted directed-worm algorithm}\label{alg:ldwa}
  \begin{algorithmic}
    \Require A bond configuration under the loop constraint  
    \State $N_{\rm w} \gets$ the total number of worms
    
    \For{$n \gets 1$ to $N_{\rm w}$}
    \State Choose a bond $b_0$, a moving direction, and a mode ($+$ or $-$) randomly.
    \State $b \gets b_0$

    \Do
    \State Choose the next bond $c$ and the next mode with a certain probability.
    \If{$b \neq c$}    
    \State Update the bond variables $n_b$ and $n_c$ and keep the moving direction forward.
    \State $b \gets c$
    \Else
    \State Flip the moving direction.    
    \EndIf
    \doWhile{$c \neq b_0$}
    \State Measure observables.
  \EndFor{}
  \State Calculate the averages of observables.
\end{algorithmic}
\end{algorithm}

We use the same type of lifting in the directed-worm framework, adding the lifted variable to the configuration space.
The worm configuration has the mode ($\sigma=+$ or $-$) in addition to the position of the kinks and the moving direction.
The lifted worm algorithm does not modify the other update processes and is described by Alg.~\ref{alg:ldwa}.
Clearly, probability optimization is crucial to the effectiveness of the lifting.

\subsection{Probability optimization}
\label{sec:po}
In the standard lifting technique, the lifted variable determines the next proposed state, and the Metropolis filter gives the acceptance probability \cite{ElciGDNGD2018}.
This scheme is simple and effective, but can we optimize the whole transition probability without decomposing the transition probability into the proposal and the acceptance probabilities?
We show here that we can easily optimize the whole probability using the geometric allocation approach \cite{SuwaT2010,TodoS2013,Suwa2014}.

\begin{figure*}
  \begin{center}
\includegraphics[width=1.9\columnwidth]{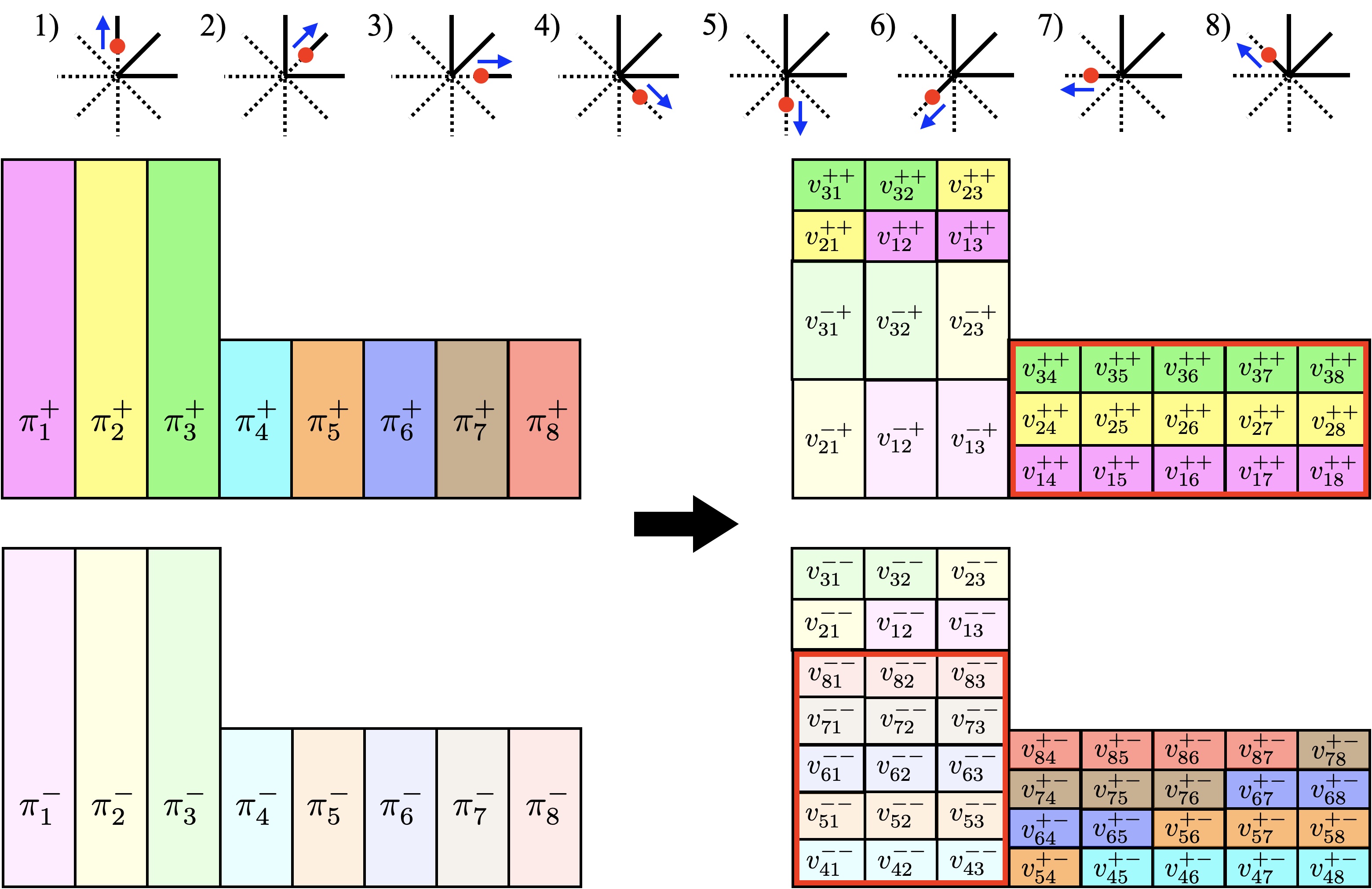}
\caption{Example of the lifted worm scattering and the geometric allocation in the four-dimensional hypercubic lattice Ising model.
In each state 1, 2, $\dots$ 8, projected onto the two-dimensional space for the illustrative purpose, the solid black lines, the red circles, and the blue arrows represent activated bonds, the moving kink, and the moving direction, respectively.
The weight of each state determined by the partition function is denoted by $\pi_i$ ($i=1,2, \dots 8$), and the weight of the lifted configuration is $\pi_i^\sigma = \frac{1}{2}\pi_i$ ($\sigma=\pm$).
$L$ and $S$ denote the sets of the states with the larger and smaller weights, respectively: $L = \{i: \pi_i > \pi_{j\in L^C} \}$ and $S=L^C$. 
The numbers of elements are $n_L=|L|$ and $n_S=|S|$, and $n_L=3$ and $n_S=5$ in this case.
Defining $\pi_L=\pi_{i\in L}$, $\pi_S=\pi_{i \in S}$, $v_a = \frac{\pi_S}{n_L}$, $v_b = \frac{\pi_L - n_S v_a}{n_L - 1}$, we set $v^{++}_{ij} = v_b \ (i,j \in L, i \neq j)$, $v_a \ (i\in L, j \in S)$, $v^{--}_{ij} = v_b \ (i,j \in L, i \neq j)$, $v_a \ (i\in S, j\in L)$, $v^{+-}_{ij} = \frac{\pi_S}{n_S - 1} \ (i,j\in S, i \neq j)$, $v^{-+}_{ij} = \frac{n_S v_a}{n_L - 1} \ (i,j \in L, i \neq j)$, and $v^{\sigma \sigma'}_{ij}=0$ for the other cases.
The areas enclosed by the red lines represent the total net stochastic flow in each mode, which is maximized in this case and amounts to $n_S \pi_S$.
The transition probability is given by $p_{(i,\sigma) \to (j,\sigma')} = v_{ij}^{\sigma \sigma'}/\pi_i^\sigma$. The order of the states within $L$ and $S$ does not matter to the transition probability.
} 
  \label{fig1}
  \end{center}
\end{figure*}

To see our idea of probability optimization, let us consider a directed-worm scattering process for the $d$-dimensional hypercubic lattice model.
Let $\pi_i$ $(i=1,...,n)$ denote the weight, or the measure, of each state, with $n=2d$.
An example of the worm scattering for $d=4$ is illustrated in Fig.~\ref{fig1}.
There are eight possible states ($n=8$) after the worm scattering.
Let $L$ and $S$ denote the sets of the states with the larger and smaller weights, respectively, that is, $L = \{i: \pi_i > \pi_{j\in L^C} \}$ and $S=L^C$.
Accordingly, let us define $\pi_L=\pi_{i\in L}$, $\pi_S=\pi_{i \in S}$, $n_L=|L|$, and $n_S=|S|$.
In the Ising model~(\ref{eq:Z}), the smaller weight states have a larger number of activated bonds $\ell=\sum_b n_b$, and the weight ratio is given by $\pi_S / \pi_L = \tanh K$. Note that $n_L$ and $n_S$ take odd numbers in the worm scattering.

We duplicate here the configurations and introduce the lifted variables.
The weight of each lifted configuration is set to half the original weight: $\pi_i^\sigma = \frac{1}{2}\pi_i$ ($\sigma=\pm$).
Our purpose is to optimize the transition probabilities from $(i,\sigma)$ to $(j,\sigma')$, $p_{(i,\sigma) \to (j,\sigma')}$ $(i,j=1,...,n$ and $\sigma,\sigma'=+,-)$, under the global balance condition:
\begin{equation}
  \pi_i^\sigma = \sum_{j,\sigma'} \pi_i^\sigma p_{(i,\sigma) \to (j,\sigma')} = \sum_{j,\sigma'} \pi_j^{\sigma'} p_{(j,\sigma') \to (i,\sigma)} \ \ \  \forall i,\sigma.
  \label{eq:gb}
\end{equation}
It is convenient to define the stochastic flow from $(i,\sigma)$ to $(j,\sigma')$: $v_{ij}^{\sigma \sigma'} \equiv \pi_i^\sigma p_{(i,\sigma) \to (j,\sigma')}$. 
The balance condition~(\ref{eq:gb}) is expressed by
\begin{equation}
    \pi_i^\sigma = \sum_{j,\sigma'} v_{ij}^{\sigma \sigma'} = \sum_{j,\sigma'} v_{ji}^{\sigma' \sigma} \quad \forall i,\sigma.
    \label{eq:gb2}
\end{equation}
The guideline of the probability optimization in our approach is to minimize the worm backscattering flow $v_{\rm back} = \sum_{i,\sigma,\sigma'} v_{ii}^{\sigma \sigma'}$ and maximize the net stochastic flow in each mode $v_{\rm net}^+ = \sum_{i\in L, j\in S} v_{ij}^{++}$ and $v_{\rm net}^-=\sum_{i \in S, j \in L} v_{ij}^{--}$.

Here, we can view the optimization problem as an allocation problem of the stochastic flow $v_{ij}^{\sigma \sigma'}$.
As illustrated in Fig.~\ref{fig1}, an allocation of $v_{ij}^{\sigma \sigma'}$ satisfies Eq.~(\ref{eq:gb2}) if each {\it color} area and the entire {\it box} shape are unchanged.
In our approach, we maximize the net stochastic flow under the condition that the backscattering probability is minimized.
We geometrically found the optimal allocation and obtained the analytical form.
For $1< n_L < n-1$, defining $v_a = \frac{\pi_S}{n_L}$ and $v_b = \frac{\pi_L - n_S v_a}{n_L - 1}$, we set 
\begin{equation}
v^{\sigma \sigma'}_{ij} =
\begin{cases}
    v_b & (\sigma=\sigma'=+, i,j \in L, i \neq j)\\
    v_a & (\sigma=\sigma'=+, i\in L, j \in S) \\ 
    v_b & (\sigma=\sigma'=-, i,j \in L, i \neq j) \\
    v_a & (\sigma=\sigma'=-, i\in S, j\in L) \\
    \frac{\pi_S}{n_S - 1} & (\sigma=+,\sigma'=-, i,j\in S, i \neq j) \\
    \frac{n_S v_a}{n_L - 1} & (\sigma=-,\sigma'=+, i,j \in L, i \neq j) \\
    0 & ({\rm otherwise}).
\end{cases}
\label{eq:v1}
\end{equation}
This solution for $n_L=3$ is shown graphically in Fig.~\ref{fig1} and is possible if $v_b \geq 0$, that is,
\begin{equation}
n_L \pi_L \geq n_S \pi_S.
\label{eq:cond1}
\end{equation}
We successfully achieve $v_{\rm back}=0$ and $v_{\rm net}^+ = v_{\rm net}^- = n_S \pi_S$, which are maximized in this case.
For $n_L=1$ and $n-1$, we avoid backscattering without net flow.
Specifically, for $n_L=1$, defining $v_c = \frac{\pi_L}{n_S}$, we set
\begin{equation}
v^{\sigma \sigma'}_{ij} =
\begin{cases}
v_c & (\sigma=\sigma', i \in L, j \in S) \\
v_c & (\sigma=\sigma', i \in S, j \in L) \\
\frac{\pi_S - v_c}{n_S - 1} & (\sigma=\sigma', i,j \in S, i \neq j)  \\
0 & ({\rm otherwise}).
\end{cases}
\label{eq:v2}
\end{equation}
This solution exists if
\begin{equation}
n_S \pi_S \geq \pi_L.
\label{eq:cond2}
\end{equation}
For $n_L=n-1$, we set
\begin{equation}
v^{\sigma \sigma'}_{ij} =
\begin{cases}
v_a & (\sigma=\sigma', i \in L, j \in S) \\
v_a & (\sigma=\sigma', i \in S, j \in L) \\
\frac{\pi_L - v_a}{n_L - 1} & (\sigma=\sigma', i,j \in L, i \neq j) \\
0 & ({\rm otherwise}).
\end{cases}
\end{equation}
This solution always exists.
Note that for $n_L=1$ and $n-1$, introducing a net stochastic flow between $L$ and $S$ necessarily causes worm backscattering.
We put backscattering minimization before stochastic flow maximization.

As in the other applications of lifting, our solution partially satisfies the skewed detailed balance \cite{TuritsynCV2011}: $\pi_i^+ p_{(i,+) \to (j,+)} = \pi_j^- p_{(j,-) \to (i,-)}$, that is, $v_{ij}^{++}=v_{ji}^{--}$.
However, in contrast to the standard lifting method, we allow the simultaneous update of both the original state variables and the lifted variable to avoid backscattering: $v_{ij}^{\sigma \sigma'} =     \frac{\pi_S}{n_S - 1} \  (\sigma=+,\sigma'=-, i,j\in S, i \neq j)$ and $\frac{n_S v_a}{n_L - 1} \  (\sigma=-,\sigma'=+, i,j \in L, i \neq j)$ in Eq.~(\ref{eq:v1}).
Although it is nontrivial to prove the ergodicity of the present algorithm mathematically because of this diagonal update, we have not observed any sign of the ergodicity breaking in our simulations.

In the Ising model~(\ref{eq:Z}), $\pi_S/ \pi_L = \tanh K$, and the condition~(\ref{eq:cond1}) is satisfied if $n_L / n_S \geq \tanh K$.
For $n_L \geq n_S$, this is always satisfied, and for $n_L < n_S$, it is satisfied if 
$K \leq \tanh^{-1} \frac{n_L}{n_S} = \frac{1}{2} \ln \frac{n}{n_S - n_L}$.
The case of $n_L=3$ requires the tightest condition.
Hence, the solution~(\ref{eq:v1}) exists for all the cases if
$K \leq \frac{1}{2} \ln \frac{n}{n-6} = \frac{1}{2} \ln \frac{d}{d-3} \Longleftrightarrow \frac{2}{\ln \frac{d}{d-3}} \leq T$.
On the other hand, the condition~(\ref{eq:cond2}) is satisfied if $1/n_S \leq \tanh K$.
Thus, the solution~(\ref{eq:v2}) exists if
$K \geq \frac{1}{2} \ln \frac{n}{n-2} = \frac{1}{2} \ln \frac{d}{d-1} \Longleftrightarrow T \leq \frac{2}{\ln \frac{d}{d-1}}=T_{\rm Bethe}$, which is the transition temperature in the Bethe approximation and always higher than the true transition temperature in finite dimensions.

In simulation, we calculate all the transition probabilities before Monte Carlo sampling and store them in memory. 
We choose the next bond and the mode for each worm scattering process using the stored probabilities and Walker's alias method \cite{FukuiT2009,HoritaST2017}. 
The computational cost of generating a random event is $O(1)$ and almost independent of the number of choices $n$ in contrast to the cost of the binary search $O(\log_2 n)$. 
The present worm algorithm has practically no extra computational cost compared to the previous worm algorithms.
We confirmed that in our implementation, the wall clock time per worm scattering of the present worm algorithm was almost the same with the P-S worm algorithm, as discussed in Appendix~\ref{wct}.

\begin{figure*}
  \begin{center}
    \includegraphics[width=1.8\columnwidth]{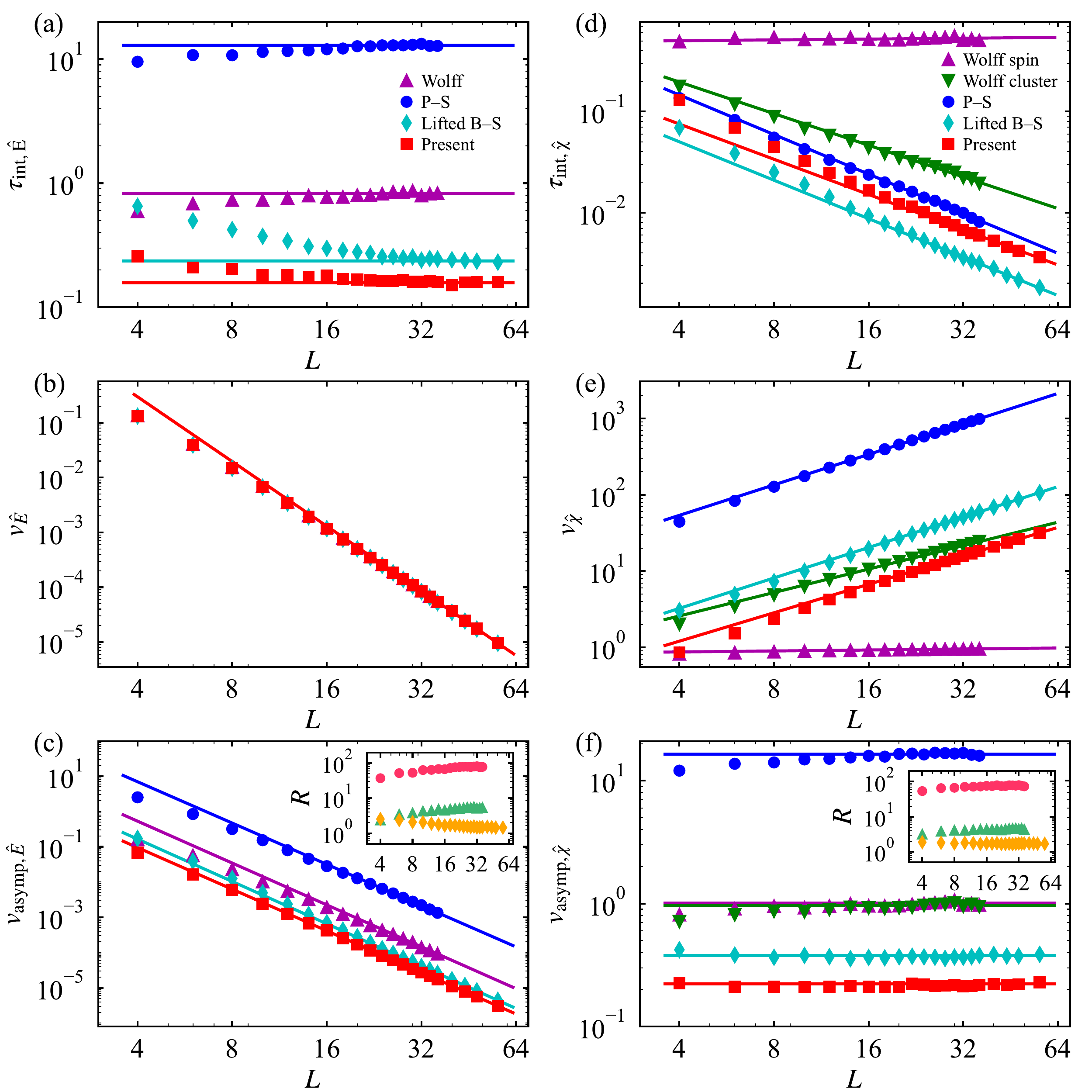}
    \caption{(a),(d) Integrated autocorrelation time; (b),(e) variance; and (c),(f) asymptotic variance of the energy estimator [(a)–(c)] and the susceptibility estimators [(d)–(f)] as a function of the system length $L$ of the four-dimensional hypercubic lattice Ising model at the critical temperature. 
    The Wolff algorithm (triangles) \cite{Wolff1989}, the Prokof'ev–Svistunov worm (circles) \cite{Prokof'evS2001}, the lifted Berretti–Sokal worm (diamonds) \cite{ElciGDNGD2018}, and the present worm (squares) algorithms are compared.
    In the Wolff algorithm, we test two susceptibility estimators using the spins dubbed the Wolff spin (upper triangles) and the cluster size dubbed the Wolff cluster (lower triangles). 
    The data are fitted to a power-law function of $L$.
    The exponents of $\tau_{{\rm int}, \hat{E}}$, $v_{\hat{E}}$, and $v_{{\rm asymp}, \hat{E}}$ are estimated to be approximately $0$, $-3.9$, and $-3.9$, respectively, and likely the same for all the four algorithms. 
    The exponents of $\tau_{{\rm int}, \hat{\chi}}$ are estimated to be approximately $0$, $-1.0$, $-1.3$, $-1.3$, and $-1.2$ for the Wolff spin, the Wolff cluster, the Prokof'ev–Svistunov worm, the lifted Berretti–Sokal worm, and the present worm algorithms, respectively; those of $v_{\hat{\chi}}$ are $0$, $1.0$, $1.3$, $1.3$, and $1.2$, respectively; that of $v_{{\rm asymp}, \hat{\chi}}$ is $0$ and presumably the same for the four algorithms. 
    The insets of panels (c) and (f) show the ratios of the asymptotic variances in the Wolff (triangles), the Prokof'ev–Svistunov worm (circles), and the lifted Berretti–Sokal (diamonds) algorithms to the one in the present worm algorithm, which are, for largest system sizes, approximately $5.4$, $80$, and $1.5$ for (c) the energy; and $4.5$, $77$, and $1.7$ for (f) the susceptibility. 
    The error bars are smaller than the symbol sizes.}
  \label{fig2}    
  \end{center}
\end{figure*}

\section{Results}
\label{sec:result}
We investigate the performance of the lifted version of the directed-worm algorithm, comparing it with the P-S worm algorithm \cite{Prokof'evS2001}, the lifted B-S worm algorithm \cite{ElciGDNGD2018}, and the Wolff cluster algorithm \cite{Wolff1989}.
We focus on the critical slowing down of the four-dimensional ($d=4$) hypercubic Ising model, and we calculate the energy and the magnetic susceptibility up to $L=56$, where $L$ is the system length, and the number of sites is $N=L^4$.
See Appendix~\ref{measurement} for the measurements of these quantities.
The temperature was set to the critical temperature $1/T_{\rm c}=0.1496947(5)$ \cite{LundowM2009}, and periodic boundaries were used in all spatial directions.
The solutions presented in Sec.~\ref{sec:po} exist because $1.443 \simeq \frac{2}{\ln \frac{d}{d-3}} \leq T_{\rm c} \leq \frac{2}{\ln \frac{d}{d-1}} \simeq 6.952$, as mentioned above.
In the present simulations, we set the initial state with $n_b=0 \ \forall b$. 
We ran $2^{28}$, $2^{24}$, $2^{24}$ and $2^{22}$ Monte Carlo steps for each Markov chain of the P-S worm, the lifted B-S worm, the present worm, and the Wolff cluster algorithms, respectively. 
The first half of the samples was discarded for thermalization (burn-in). 
The mean squared errors and the asymptotic variances of the observables were calculated using the binning analysis such that the size of each bin is much larger than the autocorrelation time \cite{Berg2004}.
We confirmed that the mean squared errors converged within the error bars with respect to the bin size. 
We also averaged the quantities over 128 independent Markov chains for each $L$, allowing us to estimate the error bars of the present results reliably. 
In the Wolff algorithm, we test two susceptibility estimators: $\hat{\chi} = \beta M_z^2 /N$ (dubbed Wolff spin), where $M_z$ is the total magnetization, and $\hat{\chi} = \beta \ell_{\rm cl} $ (Wolff cluster), where $\ell_{\rm cl} $ is the cluster size. 
To compare efficiency fairly, we normalize the autocorrelation time in units of the number of sites for each $L$, using the average worm length and cluster size.
See Appendix~\ref{comparison} for the normalization of the autocorrelation time and the definitions of the integrated autocorrelation time, the variance, and the asymptotic variance.
Note that the inverse of the asymptotic variance is proportional to the sampling efficiency of an MCMC sampler.

Figure~\ref{fig2} shows the integrated autocorrelation time $\tau_{{\rm int}, \hat{\mathcal O}}$, the variance $v_{\hat{\mathcal O}}$, and the asymptotic variance $v_{{\rm asymp}, \hat{\mathcal O}}$ of the energy ($\hat{\mathcal O}=\hat{E}$) and the susceptibility estimators ($\hat{\mathcal O}=\hat{\chi}$).
Fitting a power law to the data, we estimated the exponents, as shown in the caption. 
The exponents of the energy estimator are likely the same for the four algorithms.
Nevertheless, the sampling efficiency of the present worm algorithm is approximately $5.4$, $80$, and $1.5$ times as high as those of the Wolff cluster, the P-S worm, and the lifted B-S worm algorithms, respectively, as shown in the inset of Fig.~\ref{fig2}~(c).
Interestingly, for small system sizes, $\tau_{{\rm int}, \hat{E}}$ decreases with $L$ for the lifted B-S worm and the present algorithms while increasing for the P-S worm and the Wolff cluster algorithms.
Also, the exponent of the asymptotic variances of the susceptibility estimators is almost zero for the four algorithms. 
Again, the sampling efficiency of the present worm algorithm is approximately $4.5$, $77$, and $1.7$ times those of the Wolff cluster, the P-S worm, and the lifted B-S worm algorithms, respectively, as shown in the inset of Fig.~\ref{fig2}~(f). 
Although the size scalings of the $\tau_{{\rm int}, \hat{\chi}}$ and $v_{\hat{\chi}}$ are quite different for the two estimators in the Wolff algorithm, the asymptotic variance, that is, the product of the two quantities, as shown in Eq.~(\ref{asymp_var}), is almost the same.
Noticeably, the exponents for the P-S worm and the lifted B-S worm algorithms are almost identical.

The exponent of $v_{\hat{E}}$ should be asymptotically the same for the specific heat, $\alpha/\nu - d = -4$, where $\alpha=0$ and $\nu=1/2$ are the mean field critical exponents of the specific heat and the correlation length, respectively. 
Note that the variance is relative to the mean squared, defined by Eq.~(\ref{v_obs_est}).
The estimate obtained by a simple power law, $-3.9$, is slightly larger than the theoretical value and still decreases with $L$. 
The slow convergence of the powers of $\tau_{{\rm int}, \hat{E}}$ and $v_{\hat{E}}$ is presumably due to the logarithmic correction at the upper critical dimension \cite{Kenna2004}.

\section{Conclusions}
\label{sec:con}
We propose a lifted version of the directed-worm algorithm, adding the lifted variable ($+$ and $-$) to the original configuration.
The probability optimization in the worm scattering process is essential to efficient computation.
Using the geometric allocation approach, we minimize the worm backscattering probability and maximize net stochastic flow, as illustrated in Fig.~\ref{fig1}.
We provide the optimal probability set for the $d$-dimensional hypercubic lattice Ising model.
The backscattering probability reduces to zero, and the net stochastic flow is maximized in a wide range of temperatures, including the critical temperature. 
The entire procedure of the lifted directed-worm algorithm is described by Alg.~\ref{alg:ldwa}. 

We calculate the integrated autocorrelation time, the variance, and the asymptotic variance of the energy and the magnetic susceptibility estimators of the four-dimensional hypercubic lattice Ising model at the critical temperature. 
Several relevant exponents are the same for the Wolff cluster \cite{Wolff1989}, the P-S worm \cite{Prokof'evS2001}, the lifted B-S worm \cite{ElciGDNGD2018}, and the present worm algorithms. 
Nevertheless, the sampling efficiency of the present algorithm is approximately 5, 80, and 1.7 times those of the Wolff cluster, the P-S worm, and the lifted B-S worm algorithms, respectively, shown in Fig.~\ref{fig2}.
In our implementation, the present algorithm achieves the shortest real time needed for a certain relative error, as discussed in Appendix~\ref{wct}.
The improvement factor of the lifted B-S worm algorithm over the P-S worm algorithm is slightly different from the value reported in Ref.~\cite{ElciGDNGD2018}.
This is likely because, in the previous report, the number of activated bonds was measured not in the original physical space but in the enlarged state space, including the kinks.

We estimate the dynamic critical exponent associated with the integrated autocorrelation time of the energy to be $z_{\rm int, \hat{E}} \approx 0$ for all the compared algorithms.
For reversible cases, the integrated autocorrelation time satisfies a Li-Sokal-type bound [\cite{MadrasS2013}, Corollary 9.2.3]:  $ \tau_{{\rm int}, \hat{E}} \times L^d \geq {\rm const} \times {\rm Var}[\hat{E}]$.
This leads to $z_{\rm int, \hat{E}} \geq \alpha / \nu = 0$ for $d=4$.
Therefore, interestingly, the reversible algorithms, that is, the P-S worm and the Wolff algorithms, achieve this bound.

The autocorrelation function of the number of activated bonds has several exponential terms in $d \geq 4$, shown in Ref.~\cite{ElciGDNGD2018}.
Nevertheless, the scaling of the exponential autocorrelation time with respect to the system length is the same as that of the integrated autocorrelation time in $d=5$.
Assuming that this is the case for $d \geq 4$, equal to or greater than the upper critical dimension, we also estimate the dynamic critical exponent associated with the exponential autocorrelation time to be $z_{\rm exp} \approx 0$ from the scaling of the integrated autocorrelation time of the energy estimator, which is consistent with a previous estimate for the Wolff algorithm \cite{Tamayo1990}.
In contrast, the dynamic critical exponent of the three-dimensional Ising model in the worm and Wolff cluster updates is greater than the Li-Sokal-type bound: $z \approx 0.27 > \alpha / \nu \simeq 0.174$ \cite{Suwa2021}. 
Since the bound is achieved in $d=4$, we expect $z=0$ in $d \geq 4$.

The present approach of the lifted directed-worm algorithm can be applied to a variety of quantum models as well as classical models.
As long as the lookup table of the transition probability is prepared, the lookup cost in the simulation is almost independent of the number of states, thanks to Walker's alias method, as demonstrated in Appendix~\ref{wct}.
Although it is nontrivial to obtain an analytical solution of the transition probability in general cases, linear programming can numerically find a reasonable solution.
It is noteworthy that a numerically obtained solution may break ergodicity, so we need to check the validity of the solution by comparing it to other samplers.
As demonstrated here, the lifted directed-worm algorithm is expected to improve the sampling efficiency of the worm update for various systems.
Applications to other physical models are promising and need to be investigated in the future.

\begin{acknowledgments}
The author is grateful to Synge Todo for the fruitful discussions.
Simulations were performed using computational resources of the Supercomputer Center at the Institute for Solid State Physics, the University of Tokyo. 
This work was supported by the JSPS KAKENHI Grants-In-Aid for Scientific Research (No. 22K03508).
\end{acknowledgments}

\appendix
\section{Measurements}
\label{measurement}
From the bond representation of the Ising model~(\ref{eq:Z}), the total energy of the original spin system is given by
\begin{eqnarray}
  E &=& - \frac{\partial \ln Z}{\partial \beta} \nonumber \\
    &=& - N_b \tanh K - \left( \frac{1}{\tanh K} - \tanh K \right) \langle \ell \rangle \label{eq:e},
\end{eqnarray}
where the brackets $\langle \mathcal O \rangle$ denotes the Monte Carlo average of an observable $\mathcal O$. The number of Monte Carlo steps in our algorithm is identical to the number of worms from insertion to removal. 
In the directed-worm algorithm, a susceptibility estimator is given by
\begin{equation}
  \hat{\chi} = \frac{\beta}{4z w} \sum_{\rm scattering} f_{\rm rew}, \label{eq:chi}
\end{equation}
where $z$ is the coordination number ($z=2d$ for the $d$-dimensional hypercubic lattice), $w=\frac{1}{2}$ is the extra weight the worm carries, and
$f_{\rm rew} = \left( u + \frac{1}{u} \right) f_h$
is the reweighting factor with $u=\sqrt{\tanh K}$ after the worm scattering. 
$f_h$ takes $2/u$ if the head is on an activated bond ($n_b=1$), $2u$ on a deactivated bond ($n_b=0$), and $u+\frac{1}{u}$ on a half-activated bond ($n_b=\frac{1}{2}$). The reweighting factor $f_{\rm rew}$ is calculated after each worm scattering process and summed over the scattering processes from worm insertion to removal, as denoted by the summation in Eq.~(\ref{eq:chi}). We refer the reader to~\cite{Suwa2021} for detailed discussions of measurement in the enlarged state space.

\section{Efficiency quantification}
\label{comparison}
To assess the efficiency of an MCMC sampler, we calculate the integrated autocorrelation time, the variance, and the asymptotic variance of estimators. For a fair comparison, we normalize the number of Monte Carlo steps and the autocorrelation time in units of the number of sites $N$:
\begin{eqnarray}
  M &=& M' \frac{\langle \ell_{\rm worm} \rangle }{N},    \label{M}\\
  \tau_{{\rm int}, \hat{\mathcal O}} &=& \tau'_{{\rm int}, \hat{\mathcal O}} \frac{ \langle \ell_{\rm worm} \rangle }{N}   \label{tau_int_worm},
\end{eqnarray}
where $M'$ and $\tau'_{\rm int}$ are the original number of Monte Carlo steps and the original integrated autocorrelation time obtained from the simulations, respectively. The value of $M'$ is equal to the total number of worms during a simulation, and $\langle \ell_{\rm worm} \rangle$ is the average number of worm scattering processes from worm insertion to removal.

The integrated autocorrelation time of an estimator $\hat{\mathcal O}$ can be estimated by
\begin{equation}
  \tau_{{\rm int}, \hat{\mathcal O}}' = \frac{\sigma^2_{\hat{\mathcal O}}}{2 \bar{\sigma}^2_{\hat{\mathcal O}}},  \label{tau_int_est}
\end{equation}
where $\sigma^2_{\hat{\mathcal O}}$ is the mean squared error, that is, the square of the statistical error, calculated by binning analysis using a much larger bin size than the autocorrelation time, and $\bar{\sigma}^2_{\hat{\mathcal O}}$ is calculated without binning \cite{Berg2004}.

According to the central limit theorem, the relative error squared is asymptotically given by
\begin{equation}
\left(\frac{\sigma_{\hat{\mathcal O}}}{\mu_{\hat{\mathcal O}}}\right)^2 \approx \frac{v_{{\rm asymp}, \hat{\mathcal O}}}{M} \label{clt},
\end{equation}
where $\mu_{\hat{\mathcal O}} = \langle \hat{\mathcal O} \rangle$ is the mean of an unbiased estimator, and $v_{{\rm asymp}, \hat{\mathcal O}}$ is the asymptotic variance of the estimator.
In other words, the number of Monte Carlo steps needed to achieve a certain precision $\epsilon$ of an observable $\mathcal O$ is approximately given by $v_{{\rm asymp}, \hat{\mathcal O}} / \epsilon^2$. Thus, the sampling efficiency of an MCMC sampler can be quantified by the asymptotic variance and proportional to $v^{-1}_{{\rm asymp}, \hat{\mathcal O}}$.

The asymptotic variance is represented by
\begin{equation}
v_{{\rm asymp}, \hat{\mathcal O}} = 2 \tau_{{\rm int}, \hat{\mathcal O}} v_{\hat{\mathcal O}},   \label{asymp_var}
\end{equation}
where $v_{\hat{\mathcal O}}$ is the variance of an estimator $\hat{\mathcal O}$.
While $\tau_{{\rm int}, \hat{\mathcal O}}$ depends on MCMC updates, $v_{\hat{\mathcal O}}$ does not. Therefore, we need to consider both an efficient MCMC update and a beneficial estimator to reduce asymptotic variance.
From Eqs.~(\ref{M})-(\ref{asymp_var}), the asymptotic variance and the variance of an estimator can be estimated by
\begin{eqnarray}
      v_{{\rm asymp},\hat{\mathcal O}} &=& M' \left( \frac{\sigma_{\hat{\mathcal O}}}{\mu_{\hat{\mathcal O}}} \right)^2 \frac{ \langle \ell_{\rm worm} \rangle }{N} \label{v_asymp_est},\\
      v_{\hat{\mathcal O}} &=& M' \left( \frac{\bar{\sigma}_{\hat{\mathcal O}}}{\mu_{\hat{\mathcal O}}} \right)^2 \label{v_obs_est},
\end{eqnarray}
respectively. In the Wolff cluster algorithm, $\ell_{\rm worm}$ is replaced with the cluster size $\ell_{\rm cl}$.

\begin{figure}[tb]
  \begin{center}
    \includegraphics[width=1.0\columnwidth]{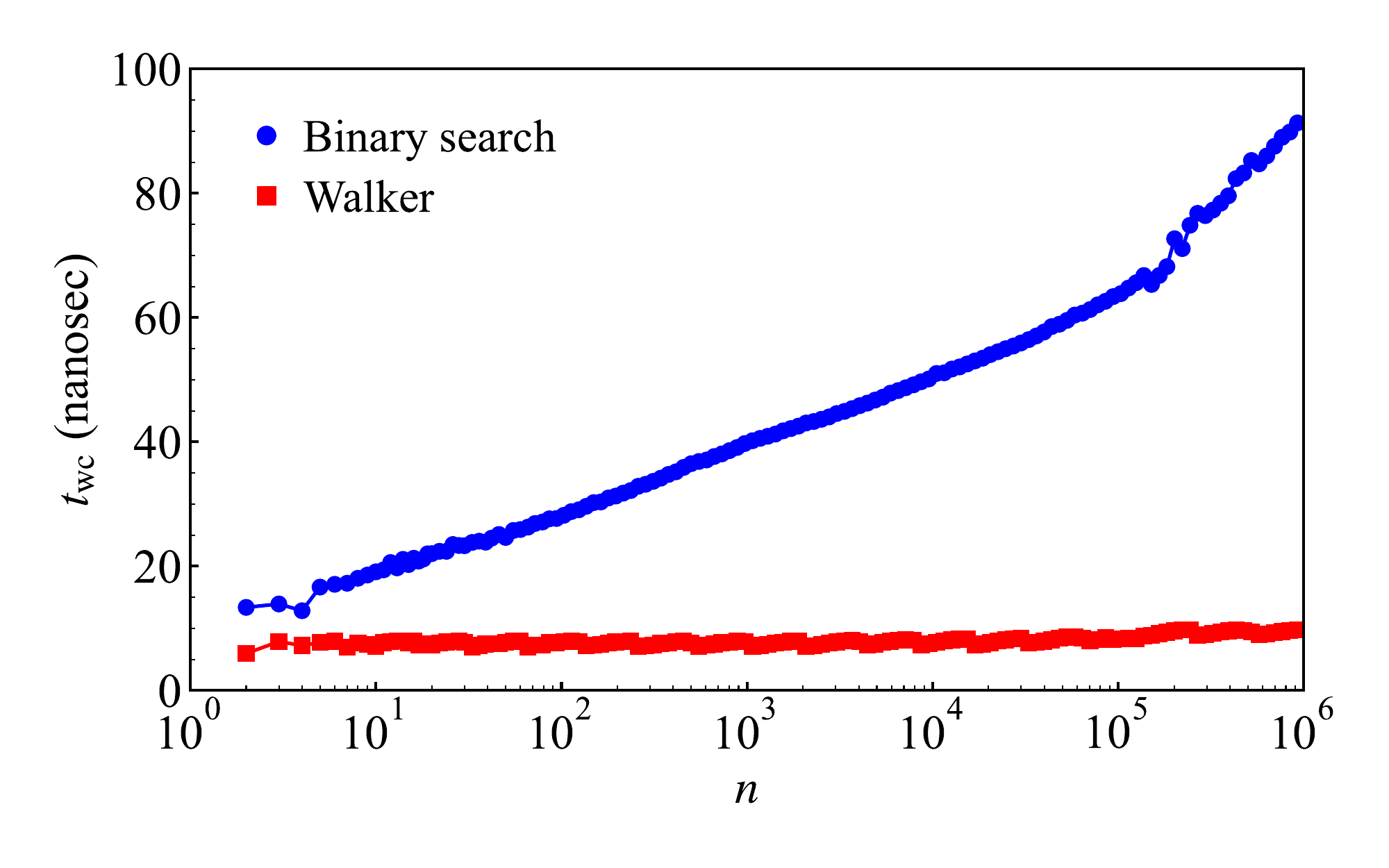}
    \caption{Wall clock time per sample of random event generation from a discrete probability distribution using the binary search and Walker's alias method.
    While $O(\log_2 n)$ in the binary search, the computation time is $O(1)$ and almost independent of $n$ in Walker's alias method.
    It increases more for $n$ larger than corresponding to the cache size, $n \sim 10^5$ in this benchmark.
    }
  \label{fig3}    
  \end{center}
\end{figure}

\section{Wall clock time}
\label{wct}
We discuss here the elapsed real time, or the wall clock time, of the present algorithm.
To assess the performance of the algorithms, we focus on the autocorrelation time and the asymptotic variance in the main text, which are independent of the implementation of the simulation code.
However, the wall clock time is also important to the actual calculation.

We measured the elapsed real time of the worm scattering in our implementation of the P-S, the lifted B-S, and the present worm algorithms on an Intel Xeon Platinum 9242 Processor using the GNU C++ compiler.
The wall clock times of the P-S and the present worm algorithms were almost the same, $\sim 100$ nanosec per scattering.
In our code, the elapsed real time of the lifted B-S algorithm was longer than that of the P-S worm algorithm, but we expect that this difference can be removed by further tuning.
We do not expect that the wall clock time of the lifted B-S algorithm can be shorter than that of the P-S worm algorithm because the P-S worm update is simple and concise.
The three algorithms thus consume almost the same real time after tuning, which is reasonable from the perspective of the shared fundamental nature of the local worm update.
From Eq.~\eqref{clt}, the real time needed to achieve a certain relative error squared is given by the product of the asymptotic variance and the wall clock time of the worm scattering process.
Therefore, the asymptotic variances shown in Figs.~\ref{fig2}(c) and \ref{fig2}(f) are practically proportional to the needed real time of the compared worm algorithms.

We also measured the wall clock time of the Wolff cluster update per site consisting of the Wolff cluster in our implementation; it was approximately three times that of the P-S and the present worm algorithms.
This longer time is due to the fact that the number of bonds checked during the Wolff cluster update is much larger than the resulting cluster size.
Thus, the needed real time of the Wolff cluster algorithm becomes larger since we normalize the number of Monte Carlo steps such that $N$ sites are updated in one Monte Carlo step, as discussed in Appendix~\ref{comparison}.
Consequently, the present algorithm also provides the most efficient sampler in real time.

We also discuss the lookup cost of the transition probability.
The number of possible states in worm scattering is $n=2d$ in the $d$-dimensional hypercubic lattice and may be quite large for other systems.
Once a lookup table of the probability is prepared, we can easily generate a random event from any discrete probability distribution.
Figure~\ref{fig3} shows the wall clock time per sample generated using the binary search and Walker's alias method \cite{FukuiT2009,HoritaST2017}.
We implemented the binary search using std::upper\_bound of the C++ Standard Library and Walker's method using the Balance Condition Library (BCL) \cite{bcl}.
The computation time of the binary search is proportional to $\log_2 n$ and increases more for $n$ larger than corresponding to the L2 cache, $n \sim 10^5$ in this benchmark.
On the other hand, the computation time of Walker's method is $O(1)$ and almost independent of $n$.
Therefore, thanks to Walker's method, the lookup cost does not depend on the model as long as the lookup table is prepared.
We also note that the computational cost of building a lookup table is $O(n)$ and is negligible compared to the Monte Carlo simulation cost \cite{FukuiT2009}.
In many cases, the present algorithm needs almost no additional computation cost compared to the P-S worm algorithm.

\bibliography{main}

\end{document}